\documentclass[intlimits,twoside,a4paper]{article}

\usepackage[cp1251]{inputenc}

\usepackage[eqsecnum]{cmpj3}

\usepackage{bm}

\issue{2020}{23}{1}{13602}
\doinumber{10.5488/CMP.23.13602}

\title[Magnetism in Rb$_{x}$Sr$_{1-x}$C]%
{Magnetism in Rb$_{x}$Sr$_{1-x}$C novel alloy}

\author[D.B. Berrezzoug, A. Lakdja]{D.B. Berrezzoug, A. Lakdja}
\address{Laboratoire de microscopie, microanalyse de la mati\`ere et spectroscopie
mol\'eculaire, Universit\'e Djillali Liab\`es de Sidi-Bel-Abb\`es 22000,
Algeria}

\date{Received June 9, 2019, in final form November 7, 2019}

\begin{document}

\maketitle

\begin{abstract}
We report a study on the electronic and magnetic properties of a novel Rb$_{x}$Sr$_{1-x}$C alloy. We find that the variation of the lattice parameter as function of Rb composition exhibits a deviation from the Vegard's law of about 0.4 {\AA}. The total energy results predict a ferromagnetic ordering for Sr-rich alloy, and antiferromagnetic ordering for RbC-rich alloy with a total magnetic moment ranging from 2 to 3$\mu_\text{B}$ per cation. As for the parent compounds, RbC and SrC, the origin of magnetism arises from the polarization of the carbon $p$-orbitals. From the spin-polarized calculations, we note that the half-metallicity in the Rb$_{x}$Sr$_{1-x}$C alloy is confirmed for $x<0.875$. On the other hand, a direct band gap is observed in the semiconducting spin part, in contrast to the pure parent compounds.
\end{abstract}

\section{Introduction}
The spintronics of semiconductors aims at combining the potential of conventional semiconductors with that of spintronics. The spintronic materials have increased the lifetime and spin diffusion length compared to conventional metals and semiconductors. In this field, diluted magnetic semiconductors (DMS) were widely studied due to their inherent properties of both semiconductor and ferromagnetic behaviour~\cite{Ohno951}.
This is the most obvious way to achieve polarization in a semiconductor and eliminate the fundamental problem of conductivity mismatch that exists between a ferromagnetic metal and a semiconductor junction~\cite{Schmidt2000}. The diluted magnetic semiconductors are composed of a semiconductor host lattice generally of classes II--VI (ZnO, CdTe, \ldots) or III--V (GaAs, InAs, \ldots) in which doping is carried out with magnetic ions such as transition metals (Mn, Co, Cr, \ldots).
In the last few years, new materials based on alkali and alkaline earth metals have been extensively studied. In this new class of magnetic materials, namely $d^{0}$-ferromagnets, the origin of magnetism arises from polarization of the anion $p$-orbitals. According to the studies undertaken in this field, these materials could be serious candidates in the spintronic field. Previous works cover binary compounds with V- and VI-elements and confirm half-metallicity with ferromagnetic ordering~\cite{Sieberer06,Yao2006,Volninanska2006,Gao2007-75,Volnianska07,Geshi07,Gao2007-91,Gao2007-19,Li08,Zhang2008-245,Zhang2008-41,Dong2011,Yan12,Zhang2012,Lakdja2013}. In particular, carbon-based compounds exhibit ferromagnetism in various structures as binaries~\cite{Gao2007-91,Zhang2008-245,Dong2011,Zhang2012,Gao2007-75} and also as heusler-type structures~\cite{Benabboun2015,Rozale14}. Moreover, elastic and vibrational stability have been confirmed for some binary compounds such as SrC and RbC~\cite{Lakdja2017}.
According to the paper of Zhang et al.~\cite{Zhang2012} and Lakdja et~al.~\cite{Lakdja2017}, the SrC compound is found to be a half-metal, whereas the RbC compound is a semiconductor. There was suggested a possibility to have half-metallicity by doping alkali compounds with alkaline earth or vice versa~\cite{Zhang2012}. This was the main source of motivation to study the Rb$_{x}$Sr$_{1-x}$C alloy. We focused on the concentration at which the alloy exhibits half-metallicity and whether it is ferromagnet (FM) or antiferromagnet (AFM). The study concerns the zinc-blende structure due to its potentiality and compatibility with other known semiconductors.

\section{Methodology}
The calculations have been performed using the plane-wave pseudopotential approach within the density functional theory (DFT)~\cite{DFT} as implemented in the Quantum-ESPRESSO package~\cite{QE}. We used the generalized gradient approximation (GGA)~\cite{GGA92} of Perdew, Burke, and Ernzerhof (PBE)~\cite{PBE} to treat the exchange-correlation potential. The projector-augmented wave (PAW) method~\cite{PAW} was used with scalar-relativistic scheme with non-linear core corrections. The valence shells considered in all calculations were as follows $4p^{6}$, $5s^{1}$ for Rb, $4p^{6}$, $5s^{2}$ for Sr, and $2s^{2}$, $2p^{2}$ for C. The plane-wave cutoff was taken up to 45~Ry, where the Gaussian smearing technique has been used. To model the Rb$_{x}$Sr$_{1-x}$C alloy with different concentrations, we built a $2\times2\times1$ 32-atoms supercell. The lattice parameter of each concentration was optimized with respect to the total energy after a full relaxation of the atomic positions within the crystal structure. The Brillouin zone (BZ) was integrated using the Monkhorste-Pack scheme~\cite{Monkhorst76} with a $3\times3\times4$ mesh.
\section{Results and discussion}
\subsection{Structural properties}
The study of the ternary alloy Rb$_{x}$Sr$_{1-x}$C was carried out by varying the composition of Rb from $x=0$ to 1. The values $x = 0$ and 1 correspond to the pure RbC and SrC, respectively. Even if the binary compounds crystallize in the rocksalt structure, they exhibit the same electronic and magnetic properties in several structures, in particular, the zinc-blende structure. The aim of this section is to see the variation of the lattice parameter of the Rb$_{x}$Sr$_{1-x}$C alloy in the zinc-blende structure according to the Vegard's law~\cite{Vegard}. The lattice parameter was calculated by optimizing the total energy for $x = 0$, 0.125, 0.25, 0.375, 0.5, 0.625, 0.75, 0.875, and 1. If we consider that bonds tend to maintain their natural length, this induces a strong individual bond distortion in the structure, and therefore a deviation from the Vegard’s law is observed. In order to estimate the degree of deviation, we fitted the calculated values with the relation;
\begin{align}
\label{vegard1}
a_{(\textrm{Rb$_{x}$Sr$_{1-x}$C})} = a_{(\textrm{Vegard})}-\delta_{a}x(1-x),
\end{align}
where
\begin{align}
\label{vegard2}
a_{(\textrm{Vegard})} = xa_{(\textrm{RbC})}+(1-x)a_{(\textrm{SrC})}
\end{align}
and $\delta_{a}$ is the deviation parameter.
\begin{figure}[!t]
\centerline{\includegraphics[scale=0.8]{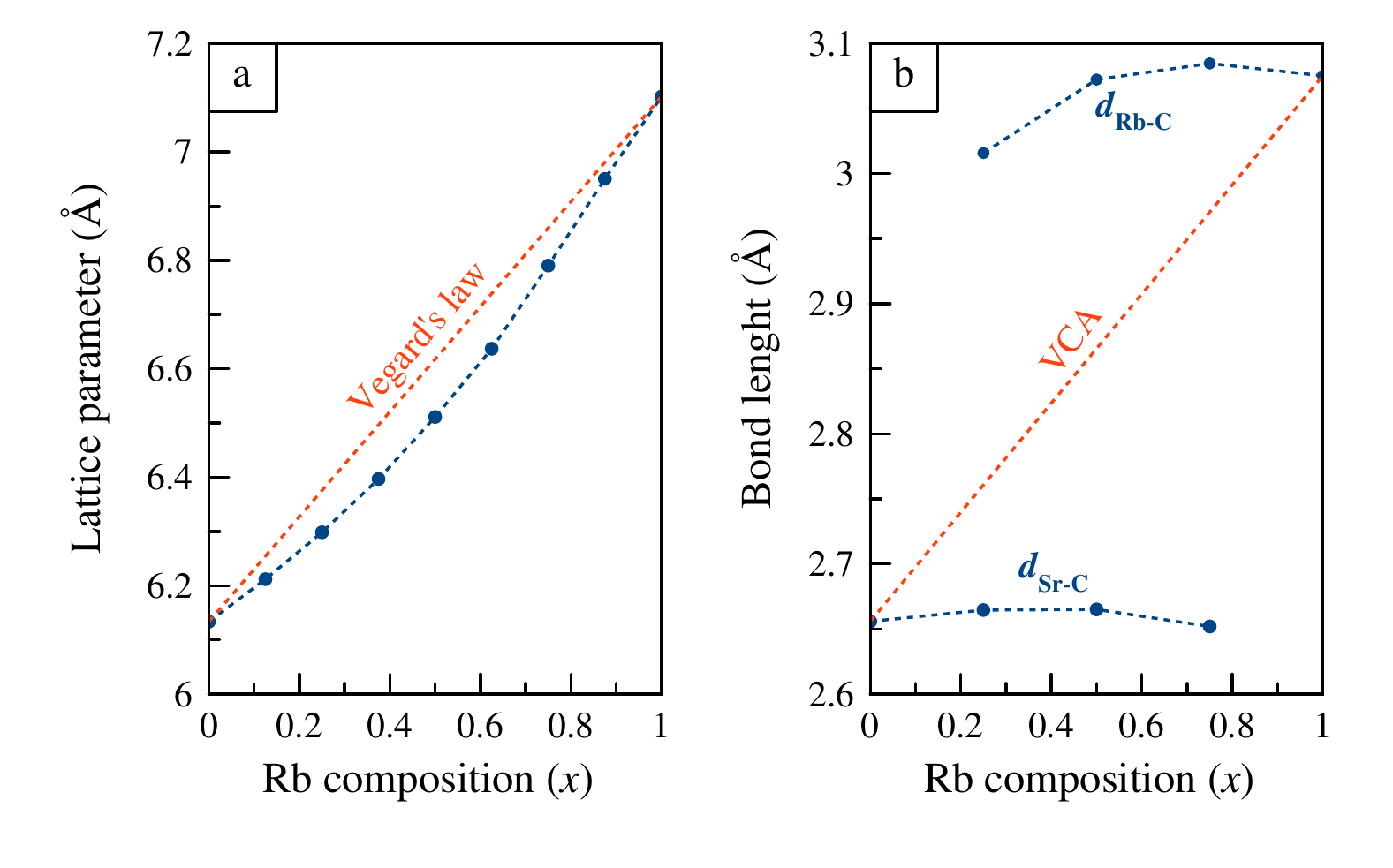}}
\vspace{-3mm}
\caption{(Colour online) a) The variation of the lattice parameter as function of Rb composition. The dotted blue curve and the dotted red line correspond to equation~(\ref{vegard1}) and (\ref{vegard2}), respectively. b) The variation of the bond lengths as function of Rb composition, the dotted red line corresponds to the VCA variation. The filled circles are the calculated values.} \label{fig1}
\end{figure}
\begin{figure}[!t]
\centerline{\includegraphics[scale=1.0]{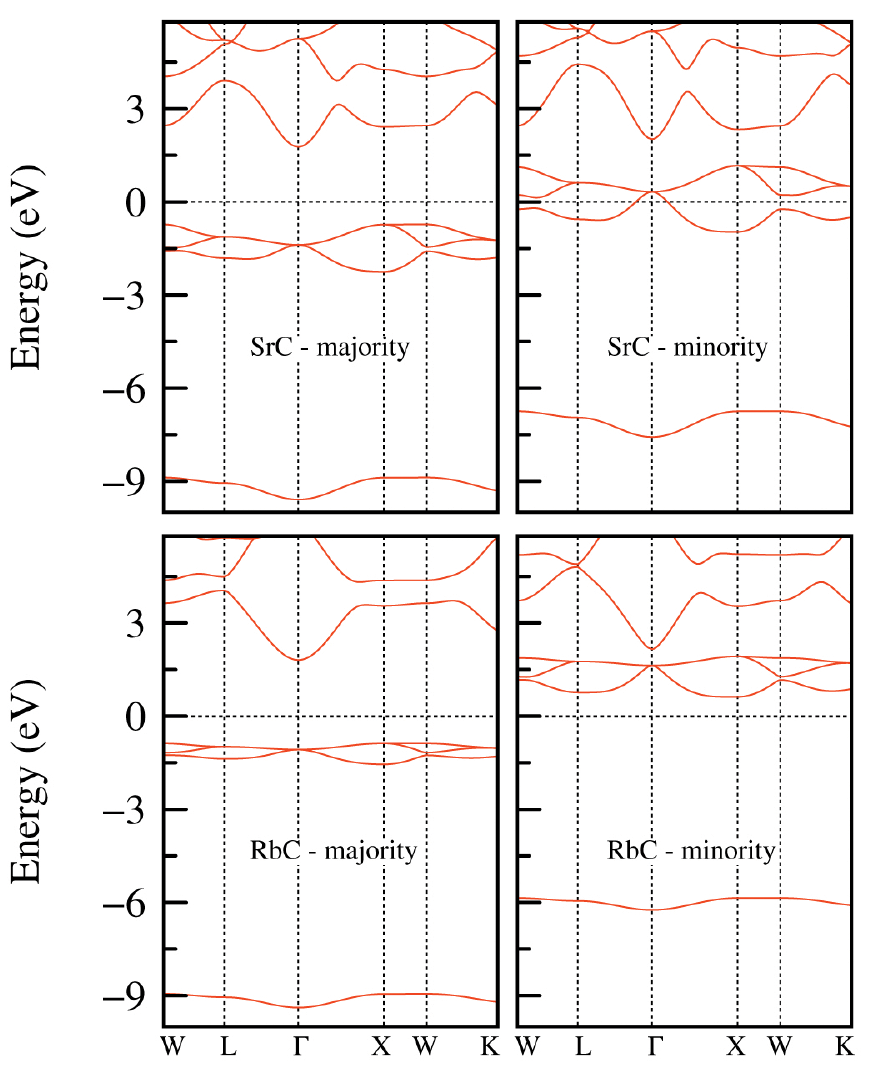}}
\caption{(Colour online) Spin-polarized band structures of the pure compounds, RbC and SrC, in the FM zinc-blende phase. The Fermi energy is set to zero.} \label{fig2}
\end{figure}

Figure \ref{fig1}~(a) shows the variation of the lattice parameter as function of Rb composition. As expected, the lattice parameter of the Rb$_{x}$Sr$_{1-x}$C alloy increases with an increasing composition $x$ going from SrC to RbC. The deviation from the Vegard's law, clearly seen in the figure, is $\delta_{a}=0.4$~{\AA}. This deviation is generally attributed to the structural relaxation effects that are not taken into account by the Vegard's law. These effects are related, among others, to the lattice mismatch between the pure parent compounds, which is found to be $\sim16$\%. Figure \ref{fig1}~(b) shows the variation of the Sr--C and Rb--C bond lengths as a function of the Rb composition. Based on the VCA approximation, the bond lengths must be identical and vary linearly with concentration $x$, as shown in figure \ref{fig1}~(b) (dotted red line). Here, we see clearly that the Rb--C distance increases with Rb composition while the Sr--C distance remains constant. Moreover, the two bond lengths are different and closer to those of the parent compounds than those predicted by the VCA. This behaviour is also related to the structural relaxation, because each carbon in the cell is surrounded by a mixture of Sr and Rb atoms, thus defining a shell with two different bonds, Sr--C and Rb--C.
\subsection{Electronic properties}
\subsubsection{The half-metallicity of SrC and RbC}

\begin{figure}[!t]
\vspace{-3mm}
\centerline{\includegraphics[scale=1.1]{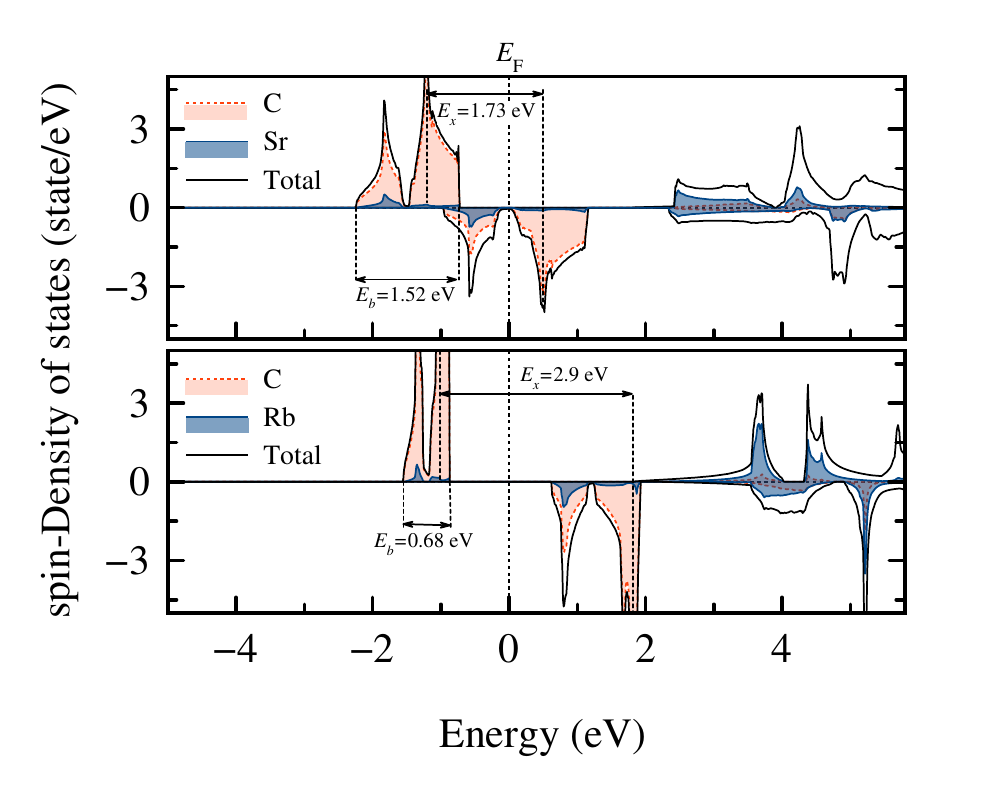}}
\vspace{-3mm}
\caption{(Colour online) Spin-dependent total and atomic density of states in the FM zinc-blende phase of the parent compounds, RbC and SrC. The Fermi energy is set to zero.} \label{fig3}
\end{figure}

\begin{table}[!b]
\caption{The lattice parameter $a$, the energy difference between the AFM and FM state $E_{\text{AFM-FM}}$, the total magnetic moment $m_\text{tot}$ for the FM state, and the magnetic ground state (MGS) at different Rb composition.}
\label{table1}
\vspace{2ex}
\begin{center}
\renewcommand{\arraystretch}{0}
\begin{tabular}{|c||c|c|c|c|}
\hline\hline
Chemical composition & $a$ (\AA) &$E_\textrm{AFM-FM}$ (meV/cation)& $m_\text{tot}$ ($\mu_\textrm{B}$/cation) & MGS\strut\\
\hline
\rule{0pt}{2pt}&&&&\\
\hline
%\raisebox{-1.7ex}[0pt][0pt]{A$_g$}
SrC                       & 6.13   & 74.3   & 2.0     & FM   \strut\\
Rb$_{0.125}$Sr$_{0.875}$C & 6.21   & 61.2   & 2.125   & FM   \strut\\
Rb$_{0.25}$Sr$_{0.75}$C   & 6.30   & 41.6   & 2.25    & FM   \strut\\
Rb$_{0.375}$Sr$_{0.625}$C & 6.40   & 27.0   & 2.375   & FM   \strut\\
Rb$_{0.5}$Sr$_{0.5}$C     & 6.51   & 12.2   & 2.5     & FM   \strut\\
Rb$_{0.625}$Sr$_{0.375}$C & 6.64   & $-$17.3  & 2.625   & AFM  \strut\\
Rb$_{0.75}$Sr$_{0.25}$C   & 6.79   & $-$21.6  & 2.75    & AFM  \strut\\
Rb$_{0.875}$Sr$_{0.125}$C & 6.95   & $-$23.4  & 2.875   & AFM  \strut\\
RbC                       & 7.10   & $-$326.0 & 3.0     & AFM  \strut\\
\hline\hline
\end{tabular}
\renewcommand{\arraystretch}{1}
\end{center}
\end{table}

In this section we investigate the half-metallicity of the pure compounds RbC and SrC in the zinc-blende. Previous studies predicted half-metallicity for SrC with a ferromagnetic (FM) ordering, while RbC is found to be a magnetic semiconductor. These properties are confirmed not only in their most stable rocksalt structure but also in several known phases, in particular zinc-blende phase. In order to better understand the magnetic state of the parent compounds in this phase, we calculated the band structures as well as the spin-density of states shown in figures \ref{fig2} and \ref{fig3}, respectively. In the band structures of SrC, the majority spin part is semiconductor-like with an indirect band gap of about 2.5~eV. The minority spin part is half-metallic, due to the bands at the Fermi level. For RbC case, the two spin parts have an energy gap, which makes it a magnetic semiconductor. Although figure \ref{fig2} corresponds to the FM order, the RbC compound is found to be AFM according to the energy difference value between the AFM and FM state indicated in table \ref{table1}. According to the band structures, the nature of the band gap in the majority spin part is found to be indirect along $\Gamma{-}X$ direction for both compounds. The spin-density of states is shown in figures \ref{fig3}. As have been already stated for rocksalt and zinc-blende structure~\cite{Gao2007-75,Gao2007-91,Zhang2008-245,Dong2011,Zhang2012}, the states at the Fermi level show  polarization of the carbon $p$-orbitals. We can see two parameters that can explain the magnetic properties related to the electronic structure, the exchange splitting $E_\text{x}$, and the bandwidth $E_\text{b}$ of the valence band maximum (VBM). If we compare these two parameters between both compounds, we see that RbC have a large exchange splitting ($E_\text{x}=2.9$ eV), with a small valence bandwidth ($E_\text{b}=0.68$ eV) compared to SrC. This may explain the half-metallicity of SrC in contrast to RbC. The bands in the majority spin part are filled in both SrC and RbC. However, in the case of SrC the minority bands at the Fermi level are partially filled with one electron which makes it half-metallic. This leads to a magnetic moment equal to the number of empty bands (2$\mu_\text{B}$). In the RbC case, the three bands in the minority spin part are empty and no band crosses the Fermi level. This situation leads to a magnetic semiconductor with a magnetic moment equal to 3$\mu_\text{B}$.

\subsubsection{The half-metallicity of the Rb$_{x}$Sr$_{1-x}$C alloy}
We focus now on the electronic and magnetic properties of the Rb$_{x}$Sr$_{1-x}$C alloy. It is interesting to show the trend that emerge when varying the Rb composition, and also determine whether the alloy is semiconductor or half-metallic. The obtained values of the energy difference $E_{\textrm{AFM-FM}}$ between the AFM and FM state are listed in table \ref{table1}. Since positive or negative value of $E_{\textrm{AFM-FM}}$ means FM or AFM state, respectively, one can see that the FM ordering is more favourable for Sr-rich alloy, while Rb-rich alloy tend to be anti-ferromagnet. The calculated values of the total magnetic moments per cation are illustrated in table \ref{table1}. As expected, it ranges between 2 and 3$\mu_\text{B}$, which corresponds to those of SrC and RbC compounds, respectively.
\begin{figure}[!b]
\centerline{\includegraphics[scale=1.0]{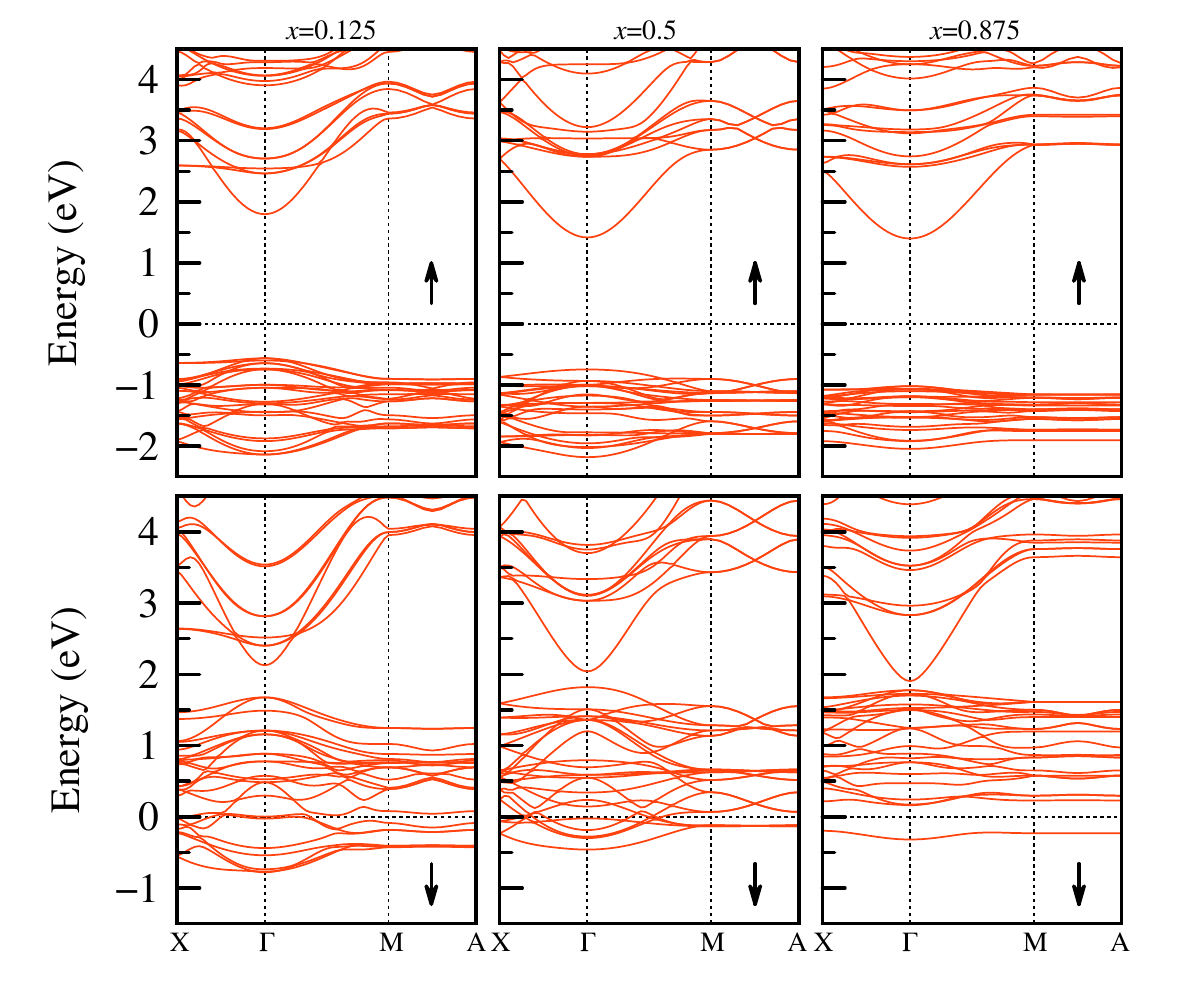}}
\caption{(Colour online) Spin-polarized band structures of the Rb$_{x}$Sr$_{1-x}$C alloy in the FM zinc-blende phase, for $x=0.125$, 0.5 and 0.875. The Fermi energy is set to zero.} \label{fig4}
\end{figure}
Figure \ref{fig4} shows the band structures for the Rb$_{x}$Sr$_{1-x}$C alloy in the zinc-blende phase for $x=0.125$, 0.5 and 0.875 (for comparison). The first notable analysis concerns the nature of the band gap. From the figure, we can see a direct transition at $\Gamma$-point, unlike the parent compounds. The same direct transition is also observed for other Rb-compositions. The second analysis is the half-metallicity observed in the minority spin part, except in the case $x=0.875$, where the band structures of the two spin parts have a semiconducting character without any band crossing the Fermi level.
\begin{figure}[!t]
\vspace{-3mm}
\centerline{\includegraphics[scale=0.95]{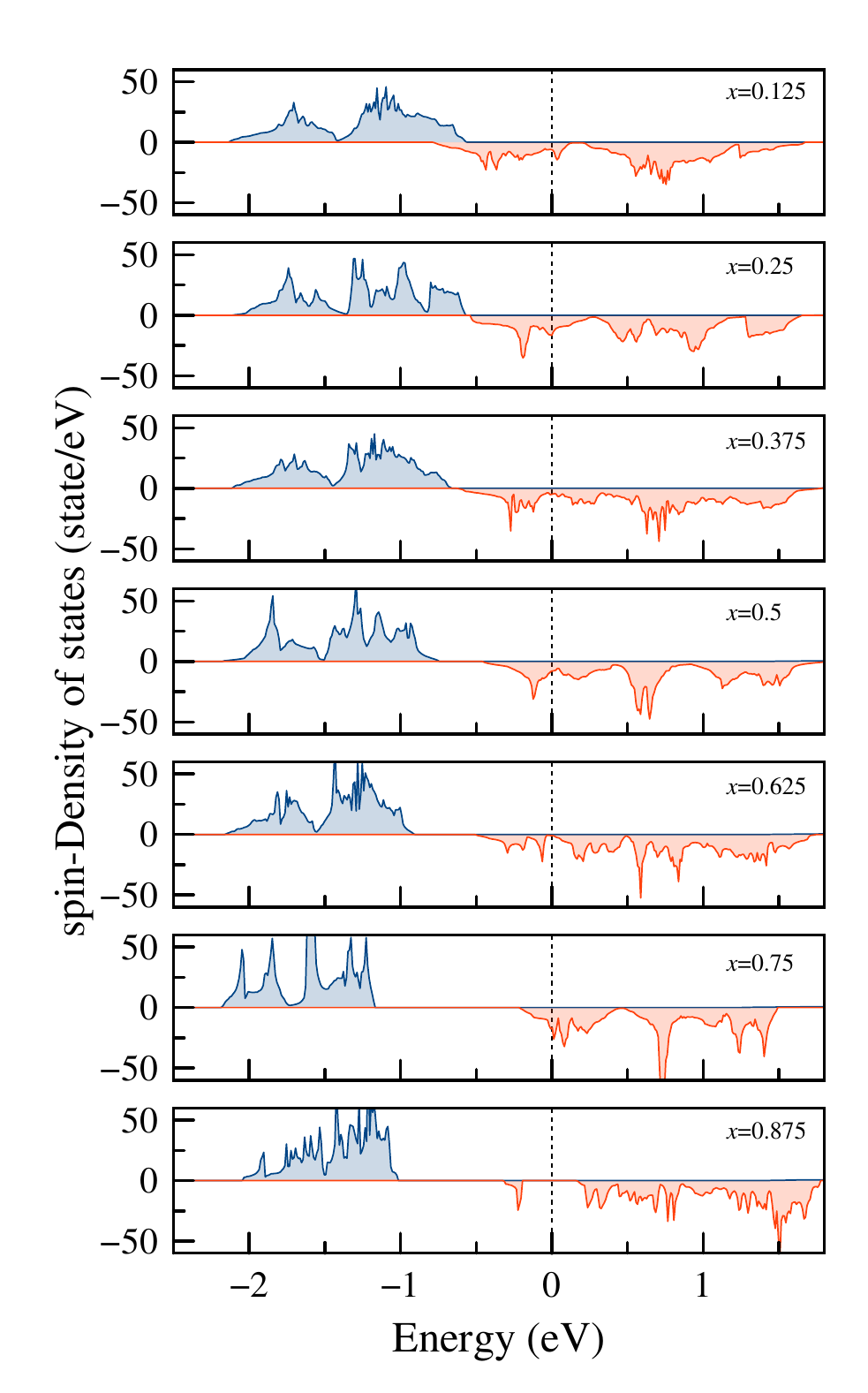}}
\vspace{-3mm}
\caption{(Colour online) Spin-dependent total density of states of the Rb$_{x}$Sr$_{1-x}$C alloy in the FM zinc-blende phase, for $x=0.125$ to 0.875. The Fermi energy is set to zero.} \label{fig5}
\end{figure}
\begin{figure}[!t]
\centerline{\includegraphics[scale=0.8]{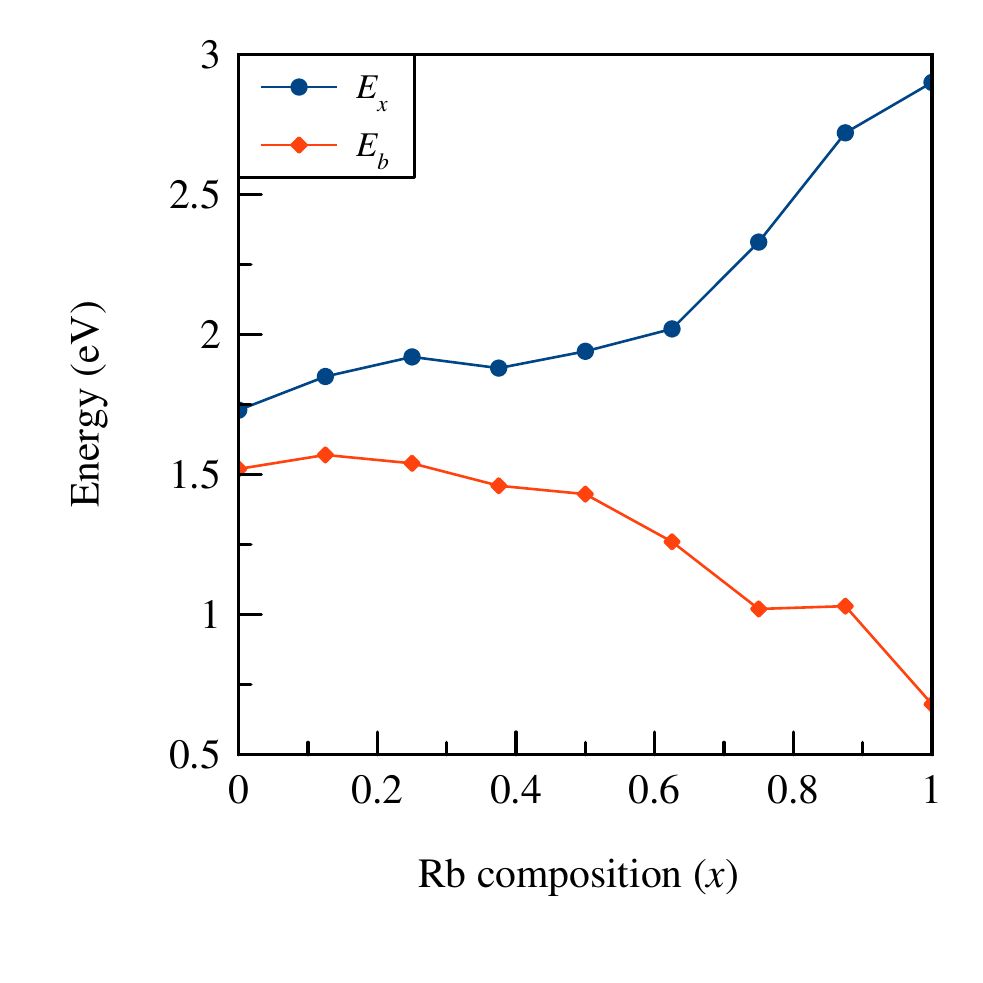}}
\vspace{-7mm}
\caption{(Colour online) The exchange splitting $E_\text{x}$ and the valence $p$-orbital bandwidth $E_\text{b}$ of the valence band maximum (VBM) as function of Rb composition. The values are in the FM zinc-blende phase of the Rb$_{x}$Sr$_{1-x}$C alloy.} \label{fig6}
\end{figure}
In order to better understand the magnetic properties of the Rb$_{x}$Sr$_{1-x}$C alloy, the spin-density of states was calculated for $x$ ranging from 0.125 to 0.875 in the FM state and represented in figure \ref{fig5}. We wanted to show only the energy region around the Fermi level which is the most interesting part. In order to better understand the trend as a function of Rb-content we calculated the values of $E_\text{x}$ and $E_\text{b}$, defined previously, of each $x$ concentration. Figure \ref{fig6} shows the variation of $E_\text{x}$ and $E_\text{b}$ as a function of the composition $x$. From these values we see that $E_\text{x}$ and $E_\text{b}$ reach a limit values at $x=0$ and 1 which correspond to those of the parent compounds. As shown in figure \ref{fig6}, $E_\text{x}$ increases with increasing $x$, while $E_\text{b}$ decreases. This induces a raise in the so-called flip between the majority and minority $p$-states. We can, therefore, associate this difference between the majority and minority $p$-orbitals to the magnetic state of the Rb$_{x}$Sr$_{1-x}$C alloy. Indeed, for small values of $E_\text{x}$, the alloy tends to a ferromagnetic material with a large $p$-band, when $E_\text{x}$ increases, the alloy becomes a semiconductor with a narrow $p$-band.

\vspace{-4mm}
\section{Conclusion}

In summary, we have used first-principles calculations to study the magnetic and electronic structures of the Rb$_{x}$Sr$_{1-x}$C novel alloy in the zinc-blende structure. The variation of the lattice parameter as function of Rb composition shows a deviation from the Vegard's law estimated to 0.4 {\AA}. The Sr-rich alloy is found to be ferromagnetic whereas RbC-rich alloy is found to be more stable in the antiferromagnetic order. A detailed analysis of the electronic structure and spin-density of states revealed a direct band gap transition at $\Gamma$-point for all concentrations. The half-metallicity is confirmed for $x$ ranging form 0 to 0.75, and may be explained from the exchange splitting $E_\text{x}$, and the bandwidth $E_\text{b}$ of the valence band maximum (VBM).

\ukrainianpart

\title{Магнетизм у новому сплаві  Rb$_{x}$Sr$_{1-x}$C }

\author{Д.Б. Беррезоу, A. Лакдья}
\address{Лабораторія мікроскопії, мікроаналізу речовини і молекулярної спектроскопії, 
університет ім. Джилалі Лябеса м. Сіді-Бель-Аббес 22000, Алжир}

\makeukrtitle

\begin{abstract}
У статті досліджуються електронні та магнітні властивості нового сплаву Rb$_{x}$Sr$_{1-x}$C. Встановлено, що зміна параметра гратки як функція складу  Rb
проявляє відхилення від закону Вегарда на величину біля  0.4 {\AA}. Результати з урахуванням сумарної енергії передбачають феромагнітне впорядкування для Sr-збагаченого 
сплаву та антиферомагнітне впорядкування для  RbC-збагаченого сплаву при сумарному магнітному моменті  в діапазоні від 2 до 3$\mu_\text{B}$ на катіон. 
Стосовно материнських сполук, RbC і SrC, встановлено, що магнетизм виникає за рахунок поляризації вуглецевих  $p$-орбіталей. Відзначається, що в результаті спін-поляризаційних обчислень напівметалічність у сплаві  Rb$_{x}$Sr$_{1-x}$C підтверджується  при  $x<0.875$. З іншого боку, на відміну від материнських сполук, заборонена зона з прямими переходами  спостерігається у напівпровідній спіновій частині.

\keywords першопринципна теорія, магнітні напівпровідники, сплави, лужні та лужно-земельні метали
\end{abstract}

\end{document}